\documentclass[9pt,conference]{IEEEtran}
\IEEEoverridecommandlockouts
\usepackage{amsmath,graphicx}
\usepackage{amssymb,amsfonts,mathtools}
\usepackage{algorithmic}
\usepackage{diagbox}
\usepackage{textcomp}
\usepackage{comment}
\usepackage{xcolor}
\usepackage{paralist}
\usepackage{symbol_maosheng}
\usepackage{multirow}
\usepackage{float}
\usepackage[caption=false, font=footnotesize]{subfig}
\usepackage{color,colortbl}
\definecolor{Gray}{gray}{0.9}
\usepackage{array}
\makeatletter
\newcommand{\thickhline}{%
    \noalign {\ifnum 0=`}\fi \hrule height 1pt
    \futurelet \reserved@a \@xhline
}
\renewcommand\baselinestretch{1}
\newcolumntype{"}{@{\hskip\tabcolsep\vrule width 1pt\hskip\tabcolsep}}
\makeatother
   
\usepackage{etoolbox}
\makeatletter
\patchcmd{\@makecaption}
  {\scshape}
  {}
  {}
  {}
\makeatother

\title{Finite Impulse Response Filters for Simplicial Complexes
\thanks{M. Yang is supported by the TU Delft AI Labs Programme. M. T. Schaub received funding from the Ministry of Culture and Science (MKW) of the German State of North Rhine-Westphalia ("NRW Rückkehrprogramm"). }}

\author{\IEEEauthorblockN{Maosheng Yang}
\IEEEauthorblockA{\textit{Dept. of Intelligent Systems} \\
\textit{Delft University of Technology}\\
Delft, The Netherlands \\
m.yang-2@tudelft.nl}
\and
\IEEEauthorblockN{Elvin Isufi}
\IEEEauthorblockA{\textit{Dept. of Intelligent Systems} \\
\textit{Delft University of Technology}\\
Delft, The Netherlands  \\
e.isufi-1@tudelft.nl}
\and
\IEEEauthorblockN{Michael T. Schaub}
\IEEEauthorblockA{\textit{Dept. of Computer Science} \\
\textit{RWTH Aachen University}\\
Aachen, Germany \\
schaub@cs.rwth-aachen.de}
\and
\IEEEauthorblockN{Geert Leus}
\IEEEauthorblockA{\textit{Dept. of Microelectronics} \\
\textit{Delft University of Technology}\\
Delft, The Netherlands\\
g.j.t.leus@tudelft.nl}
}

\begin{document}
\maketitle
\begin{abstract}
In this paper, we study linear filters to process signals defined on simplicial complexes, i.e., signals defined on nodes, edges, triangles, etc. of a simplicial complex, thereby generalizing filtering operations for graph signals. We propose a finite impulse response filter based on the Hodge Laplacian, and demonstrate how this filter can be designed to amplify or attenuate certain spectral components of simplicial signals. Specifically, we discuss how, unlike in the case of node signals, the Fourier transform in the context of edge signals can be understood in terms of two orthogonal subspaces corresponding to the gradient-flow signals and curl-flow signals arising from the Hodge decomposition. By assigning different filter coefficients to the associated terms of the Hodge Laplacian, we develop a subspace-varying filter which enables more nuanced control over these signal types. 
Numerical experiments are conducted to show the potential of simplicial filters for sub-component extraction, denoising and model approximation.
\end{abstract}

\section{Introduction}
Graph signal processing (GSP) has emerged over the last years as a powerful tool to deal with high-dimensional signals defined on non-Euclidean domains~\cite{shuman2013emerging, ortega2018graph}. 
% In GSP signals are treated as a mapping from the node set of a graph to the set of real numbers. 
Using the perspective of GSP, notions like the Fourier transform, filters, and signal sampling have been extended to the domain of graphs \cite{sandryhaila2013discrete, sandryhaila2014discrete, chen2015discrete}. 
These definitions can also be leveraged to understand graph neural networks \cite{gama2020}. 
% GSP not only provides a generalization of signal processing for time-series and images, but also providing us an underlying model, i.e., graph, to model interactions between signals.
GSP not only provides a generalization of signal processing for time series and images, but specifies an explicit model for the underlying signal dependencies in terms of a graph.

% However, the graph is limited to model complex networked data since it can only captures pairwise relationships between data entities by its edges. In a coauthorship network, for instance, we need to describe collaborations between more than two authors \cite{huang2015metrics}. 
By construction, graphs model dependencies between node data via the edges, which encode pairwise relations between nodes.
In certain scenarios, however, we may be interested in specifying multi-way relationships between subsets of nodes.
For instance, in a coauthorship network, we want to describe collaborations between more than two authors \cite{huang2015metrics}. 
% In a coauthorship network, for instance, we need to describe collaborations between more than two authors \cite{huang2015metrics}. 
Similarly, in a social network, we want to capture the connections between groups of friends rather than two people \cite{wilkerson2013simplifying}. 
To deal with such settings, researchers have introduced \emph{hypergraph signal processing} as one paradigm, in which we model group relationships through hyperedges \cite{barbarossa2016introduction, zhang2019introducing, schaub2021}. 
Other works \cite{coutino2020self, leus2021} use the Volterra model to describe higher-order relations.

In parallel, \emph{topological signal processing} (TSP) has been proposed to analyze signals defined over topological spaces, especially in the form of simplicial complexes composed of nodes, edges, and triangles, etc.~\cite{schaub2021, barbarossa2020, lim2015hodge}. 
Important examples of such signals defined on simplices include flow signals over edges, like traffic flows in a transportation network or data flows in a communication network. 
Edge flows have also been considered in the context of statistical ranking or to process currency exchange market rates \cite{jiang2011statistical}. 
To process such flow signals, a number of procedures have been developed. For instance, \cite{schaub2021,schaub2018flow} consider the problem of flow denoising by solving a regularized optimization problem, which is equivalent to a low-pass filter in the edge space. Similarly, flow interpolation and edge sampling have been studied in \cite{jia2019graph}, and random walks have been extended from the node space to the edge space in \cite{schaub2020random}. 
There exist even certain neural network architectures for data defined on the edge space of a simplicial complex \cite{roddenberry2019hodgenet}. 
However, more explicit treatments of linear filtering and filter design on simplicial complexes, have so far not been provided in the literature.
In this paper, we thus define, design and analyze finite impulse response (FIR) filters for simplicial complexes, as fundamental building blocks for signal processing on simplicial complexes.

\vspace{1mm}\noindent\textbf{Contribution and paper outline.} 
We re-examine the Fourier transform for simplicial complexes (SCs) \cite{schaub2021,barbarossa2020,schaub2018flow}, and define two types of simplicial frequencies, measuring different properties of signals. 
Using this perspective we propose an FIR filter for SCs, which employs the 1-Hodge Laplacian associated to the SC as a shift operator, in lieu of the graph Laplacian used analogously for graph signals.
Focusing on the processing of edge flows, we study this local shift operator, which enables to implement convolutions of flow on SCs locally with a small computational complexity. 
Using the Hodge decomposition associated to the space of edge flows, we show the capability of such filters to modulate signals at different frequencies. 
Further, by assigning individual filter coefficients to the lower and upper adjacent coupling terms of the Hodge Laplacian, we show how to create a more expressive linear filter that can modulate gradient and curl flows more precisely, leading to an improved filter performance. 
We illustrate our results by several numerical experiments.

The remainder of the paper is structured as follows.
In Section~\ref{sec:background} we review simplicial complexes, simplicial signals, the Hodge Laplacian and the Hodge decomposition. In Section~\ref{sec:fourier transform}, we recall the Fourier transform of simplicial signals and define two types of simplicial frequencies. The FIR filter on SCs is proposed in Section~\ref{sec:simplicial filter} where simplicial signal shifting, spectral analysis and filter design are investigated. 
Then the subspace-varying filter is proposed with improved performance. 
We conduct experiments in synthetic and real world networks in Section~\ref{sec:experiments} before concluding the paper.

\section{Background} \label{sec:background}
\noindent\textbf{Simplicial complexes and signals.}
Given a finite set of vertices $\ccalV$, a $k$-simplex $\ccalS^k$ is a subset of $\ccalV$ with cardinality $k+1$. 
A subset of a $k$-simplex with cardinality $k$ is called a \emph{face}. The number of faces of a $k$-simplex is $k+1$. A \emph{coface} of a $k$-simplex is a $(k+1)$-simplex that includes this $k$-simplex. 
A node is a $0$-simplex, an edge is a $1$-simplex, and a triangle is a $2$-simplex. For an edge, its incident nodes are faces. The edges connecting a node with its neighbors, are the cofaces of that node \cite{barbarossa2020, lim2015hodge}.  

\begin{figure}[t] 
\vspace{-7mm}
  \centering
  \subfloat[An all-one flow\label{1a}]{%
       \includegraphics[width=0.33\linewidth]{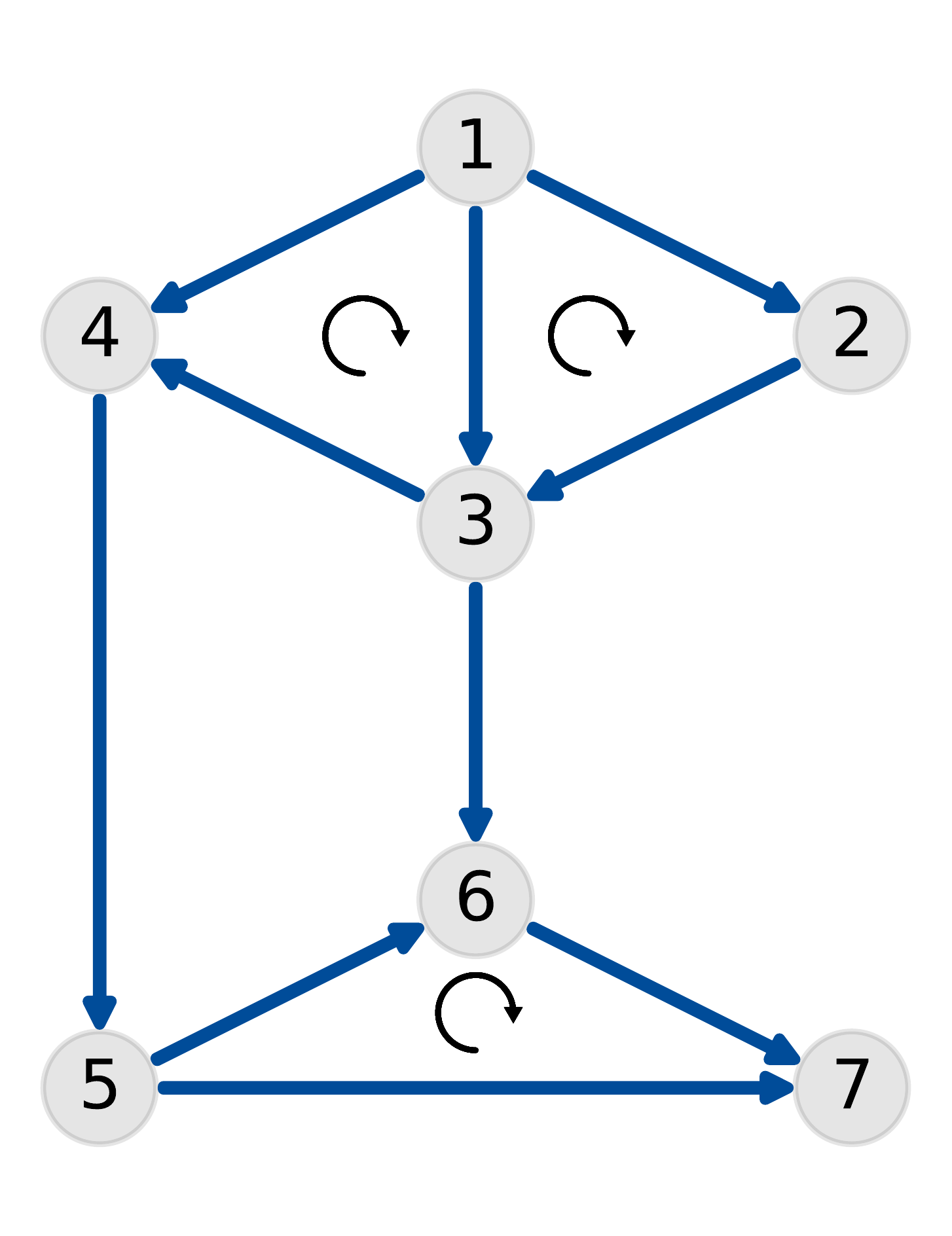}}
  \subfloat[Shifting by $\bbL_{1,\ell}$\label{1b}]{%
        \includegraphics[width=0.33\linewidth]{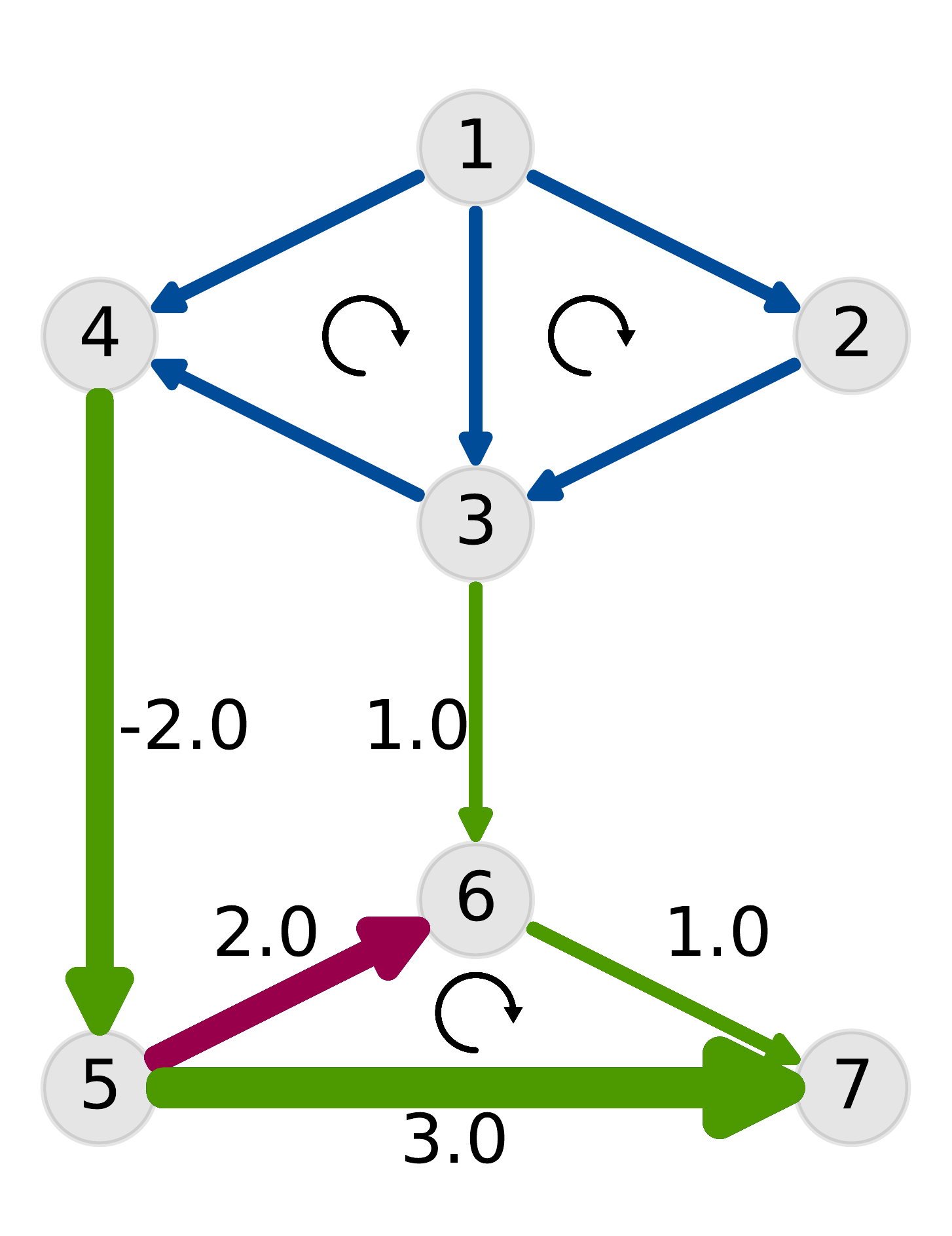}}
  \subfloat[Shifting by $\bbL_{1,u}$\label{1c}]{%
        \includegraphics[width=0.33\linewidth]{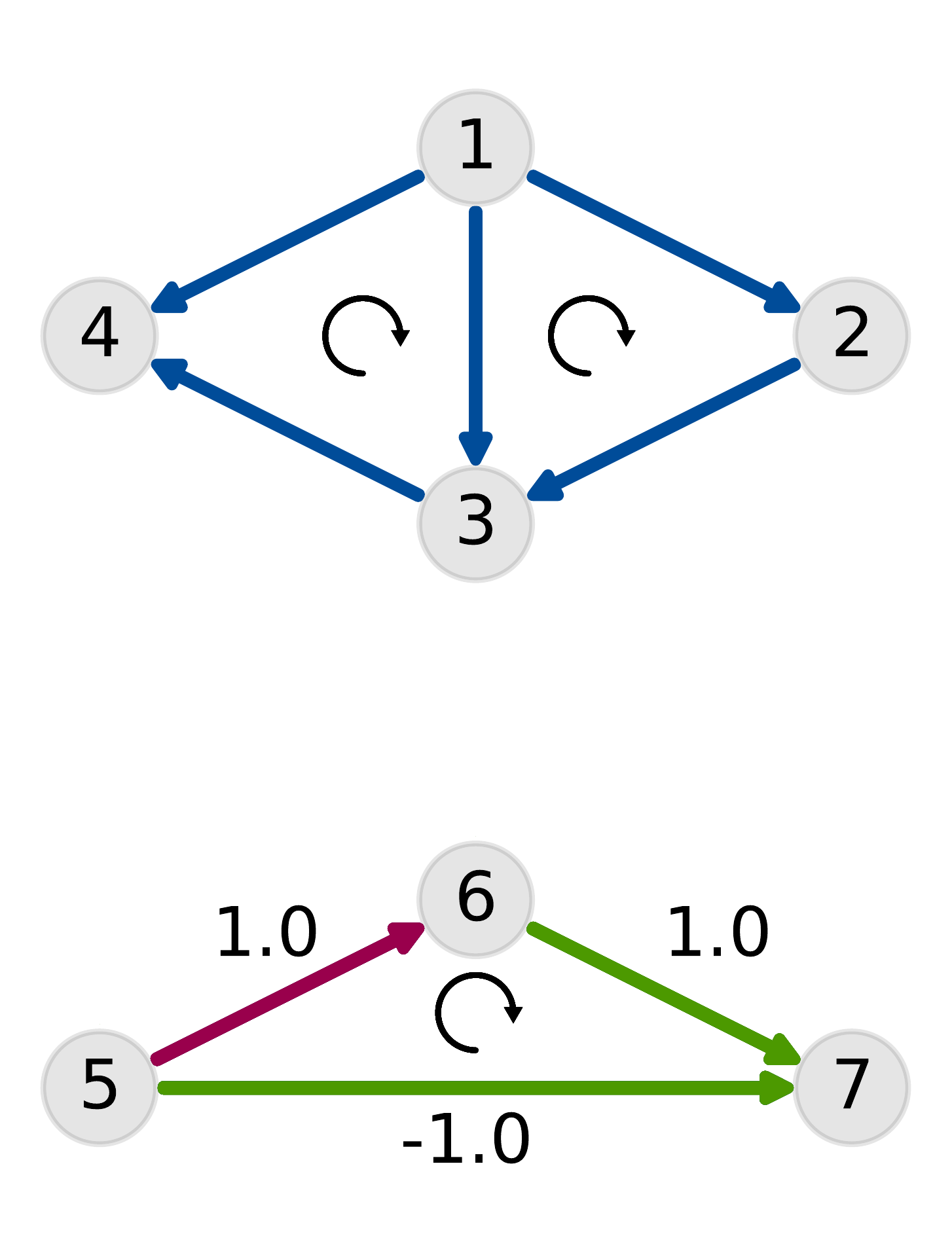}}
  \caption{(a) An all-one flow defined on the edges of a simplicial complex (SC) with three $2$-simplices $\{1,2,3\}, \{1,3,4\}, \{5,6,7\}$. The flow magnitude is displayed by the edge width.  (b-c) One-step $\bbL_{1,\ell}$ and $\bbL_{1,u}$ shifted results shown on edges only in red and green. 
  A negative value indicates an opposite relative direction and no link means zero flow on that edge. (b) In the shifting by $\bbL_{1,\ell}$, edge $(5,6)$ (in red) locally collects the information from its one-hop lower neighbors (in green) and combines with its own state. (c) In the shifting by $\bbL_{1,u}$, edge $(5,6)$ (in red) locally collects the information from its one-hop upper neighbors (in green) and combines with its own state. Thus, by summing (b) and (c), the one-step $\bbL_1$ shifted result on edge $(5,6)$ is 3.
  }
  \label{fig:1} 
  \vspace{-4mm}
\end{figure}

A simplicial complex $\ccalX$ is a set of simplices such that for any $k$-simplex $\ccalS^k$ in $\ccalX$, any subset of $\ccalS^k$ must also be in $\ccalX$. We denote the number of $k$-simplices in $\ccalX$ by $N_k$.
A graph is a simplicial complex with simplices of maximal cardinality $2$. 
Another example is shown in Fig.~\ref{1a} where nodes are the $0$-simplices, edges are the $1$-simplices, and triangles are the $2$-simplices \cite{schaub2021}. 
Note that every simplex is equipped with a reference orientation, as indicated by the arrows in Fig.~\ref{1a}.

In a simplicial complex, we define a $k$-simplicial signal as a mapping from the $k$-simplices to the real space $\setR^{N_k}$.
For example, $\setR^{N_0}$ is the graph signal space in GSP, and $\setR^{N_1}$ is the space of edge flows. 
For an edge flow $\bbf = [f_1,\dots,f_{N_1}]^\top \in\setR^{N_1}$, the sign of its entry denotes the  direction of the flow \cite{schaub2021,lim2015hodge}, relative to a chosen (arbitrary) reference orientation. For example, the direction of a random flow is shown by the edge arrows in Fig.~\ref{1b}.  

\vspace{1mm}\noindent\textbf{Incidence matrices for simplicial complexes.} 
The relationships between $(k-1)$- and $k$-simplices can be described via the incidence matrix $\bbB_k$. 
The rows of $\bbB_k$ are indexed by $(k-1)$-simplices and the columns by $k$-simplices.
The matrix $\bbB_1$ is simply the node-edge incidence matrix, and $\bbB_2$ is the edge-triangle incidence matrix.
We construct $\bbB_2$ as follows: if an edge is oriented along with its coface, then the corresponding entry in $\bbB_2$ is $1$; if it is anti-aligned, the entry is $-1$; otherwise it is zero. 
For example, in Fig. \ref{1a}, edge $\{1,2\}$ is aligned with triangle $\{1,2,3\}$, while edge $\{1,3\}$ is not.

\vspace{1mm}\noindent\textbf{The Hodge Laplacian and the Hodge decomposition.}
%\mts{added another heading here..}
An algebraic tool to analyze simplicial signals is the \emph{$k$-th Hodge Laplacian}, given by $\bbL_k = \bbB_k^\top\bbB_k + \bbB_{k+1}\bbB_{k+1}^\top$, where we define the lower Laplacian $\bbL_{k,l}\triangleq \bbB_k^\top \bbB_k$, and the upper Laplacian $\bbL_{k,u}\triangleq \bbB_{k+1} \bbB_{k+1}^\top$. 
The $0$-th Hodge Laplacian corresponds to the graph Laplacian $\bbL_0 = \bbB_1 \bbB_1^\top$ used in GSP. 
While we consider unweighted Hodge Laplacians in this paper, weighted variants also exist \cite{schaub2020random}.

Hodge Laplacians admit a \emph{Hodge decomposition}, by which the simplicial signal space can be decomposed into three orthogonal subspaces
$ 
    \setR^{N_k} = \im(\bbB_k^\top) \oplus \im(\bbB_{k+1}) \oplus \ker(\bbL_k),
$
where $\oplus$ is the direct sum of vector spaces and $\im(\cdot)$ and $\ker(\cdot)$ are the \emph{image} and \emph{kernel}
spaces of a matrix. For $k=1$, these subspaces carry the following interpretations \cite{barbarossa2020,schaub2020random}.

{\it Gradient space}. The incidence matrix $\bbB_1$ acts as the divergence operator mapping an edge flow $\bbf$ into a node signal $\bbB_1 \bbf$, which computes the net flow of each node. Its adjoint $\bbB_1^\top$ can differentiate a node signal $\bbv$ along the edges to induce an edge flow $\bbB_1^\top\bbv$, i.e., the gradient operation. Thus, any flow $\bbf_G\in\im(\bbB_1^\top)$ can be written as the gradient of a node signal $\bbv$, i.e., $\bbf_G = \bbB_1^\top\bbv$. We call $\bbf_G$ a \emph{gradient flow}. The subspace $\im(\bbB_1^\top)$ is defined as the \textit{gradient space}. 

{\it Curl space}. The incidence matrix $\bbB_2$ can induce an edge flow from a triangle signal $\bbt$ by $\bbB_2\bbt$. 
Its adjoint $\bbB_2^\top$ is known as the {\it curl operator}. By applying it to an edge flow $\bbf$ as $\bbB_2^\top\bbf$, we can compute the flow circulating along the triangles. Thus, any flow $\bbf_C \in \im(\bbB_2)$ can be induced by a triangle flow $\bbt$ as $\bbf_C = \bbB_2\bbt$. We call $\bbf_C$ a \emph{curl flow}. The subspace $\im(\bbB_2)$ is defined as the \textit{curl space}.

{\it Harmonic space}. The remaining space $\ker(\bbL_1)$ is known as the {\it harmonic space}. Any flow $\bbf_H\in\ker(\bbL_1)$ is divergence and curl free, i.e., it is flow conservative. That is, the net flow at every node is zero and the flow circulating along the triangles is also zero.

Since $\bbB_1 \bbB_2 = \bf 0$, any gradient flow $\bbf_G$ satisfies $\bbB_2^\top\bbf_G= \bf 0$, i.e., it does not include any circular flows along  triangles and is accordingly called \textit{curl-free}. 
The space orthogonal to the gradient space, i.e., $\ker(\bbB_1) = \im(\bbB_2) \oplus \ker(\bbL_1)$, is called the \textit{cycle space}. 
Any flow $\bbf$ in this space fulfills $\bbB_1\bbf= \bf0$, and is thus called \textit{divergence-free}. Note that the cycle space consists of both the curl space and the harmonic space discussed above.

\section{Simplicial Fourier transform} \label{sec:fourier transform}
In this section, we recall the Fourier transform of simplicial signals \cite{schaub2021,barbarossa2020} and define two types of simplicial frequencies.

\vspace{1mm}\noindent\textbf{Simplicial Fourier transform.} The $k$-th Hodge Laplacian has the spectral decomposition $\bbL_k = \bbU_k \bLambda_k \bbU_k^\top$, where the matrix $\bbU_k = [\bbu_{k,1},\dots,\bbu_{k,N_k}]$ collects the eigenvectors and the diagonal matrix $\bLambda_k = \diag(\lambda_{k,1},\dots,\lambda_{k,N_k})$ the corresponding eigenvalues.
Given a signal $\bbx^k\in\setR^{N_k}$ defined on the $k$-simplices, its embedding by the simplicial Fourier Transform (SFT) is $\tilde{\bbx}^k = \bbU_k^\top \bbx^k$, i.e., the projection of $\bbx^k$ on $\bbU_k$. 
The inverse SFT is $\bbx^k = \bbU_k \tilde{\bbx}^k$ \cite{barbarossa2020}. The STF of $\bbL_0$ corresponds to the graph Fourier transform \cite{sandryhaila2013discrete}.

The eigenvectors of $\bbL_1$ span the three subspaces that appeared in the Hodge decomposition: (i) the gradient space $\im(\bbB_1^\top)$ is spanned by the eigenvectors $\bbU_G$ of $\bbL_{1,\ell}$ with positive eigenvalue; (ii) the curl space $\im(\bbB_2)$ is spanned by the eigenvectors $\bbU_C$ of $\bbL_{1,u}$ with positive eigenvalue; and (iii) the harmonic space $\ker(\bbL_1)= \ker(\bbL_{1,\ell})\cap\ker(\bbL_{1,u})$ is spanned by the eigenvectors $\bbU_H$ of $\bbL_1$ with zero eigenvalue.
Moreover, the gradient and curl space span the image of $\bbL_1$, i.e., $\bbU_G \oplus \bbU_C = \im(\bbL_1)$. 
This correspondence between eigenvectors of the Hodge Laplacian and the Hodge decomposition exists in general. However, we focus on the edge space in this paper. 

Finally, since the eigenvalues of $\bbL_k$ are all non-negative they can be naturally interpreted in terms of a  frequency.
However, unlike in the case of graphs~\cite{sandryhaila2013discrete} there are two types of eigenvectors for $\bbL_k$. For $k=1$, we thus distinguish the following types of frequencies:

\textit{Gradient frequencies.} For an eigenvector $\bbu_G$ of $\bbL_1$ in the gradient space $\text{span}(\bbU_G$), the corresponding eigenvalue is $\lambda_G = \bbu_G^\top\bbL_1\bbu_G = \lVert \bbB_1 \bbu_G \lVert^2_2$, which is an $\ell_2$-norm divergence measure for $\bbu_G$. Thus, the eigenvectors in $\bbU_G$ corresponding to a large eigenvalue $\lambda_G$ have a large quadratic measure of the divergence. If the SFT of an edge flow has a large projection on such eigenvectors, we say it has a high gradient frequency. We call the eigenvalues $\lambda_G$ associated to the gradient space $\bbU_G$ \emph{gradient frequencies}.

\textit{Curl frequencies.} For an eigenvector $\bbu_C$ in the curl space $\text{span}(\bbU_C)$, its corresponding eigenvalue can be written as $\lambda_C = \bbu_C^\top \bbL_1 \bbu_C = \lVert \bbB_2^\top \bbu_C \lVert^2_2$, which is an $\ell_2$-norm curl measure for $\bbu_C$. Thus, eigenvectors corresponding to a large $\lambda_C$ have a large curl quadratic measure. 
We call the eigenvalues $\lambda_C$ associated to the curl space \emph{curl frequencies}.

\textit{Harmonic frequencies.}
Finally, we call the zero eigenvalues \emph{harmonic frequencies}. The multiplicity of zero frequencies is equal to the number of 1-dim holes (cycles) in a SC \cite{lim2015hodge}. 
If an edge flow is contained in the harmonic space, we say it is a harmonic flow. 

For convenience, we collect gradient frequencies in the set $\ccalQ_G = \{\lambda_{G,1},\dots,\lambda_{G,N_G}\}$, curl frequencies in the set $\ccalQ_C = \{\lambda_{C,1},\dots,\lambda_{C,N_C}\}$ and harmonic frequencies (which are just zeros) in the set $\ccalQ_H$.
Thus, given any edge flow $\bbf\in\setR^{N_1}$, we can use the SFT to define three embeddings by projecting $\bbf$ onto each of the subspaces defined by the Hodge decomposition:
\begin{inparaitem}
    \item[i)] the gradient embedding: $\tilde{\bbf}_G = \bbU_G^\top \bbf$;
    \item[ii)] the curl embedding: $\tilde{\bbf}_C = \bbU_C^\top \bbf$;
    \item[iii)] the harmonic embedding: $\tilde{\bbf}_H = \bbU_H^\top \bbf$.
\end{inparaitem}
If we can control these embeddings for any edge flow, we can achieve the desired \emph{filtering}. 

\begin{figure}[!t] 
  \vspace{-7mm}
  \centering
  \subfloat[Noise-free flow $\bbf_0$\label{2a}]{%
       \includegraphics[width=0.33\linewidth]{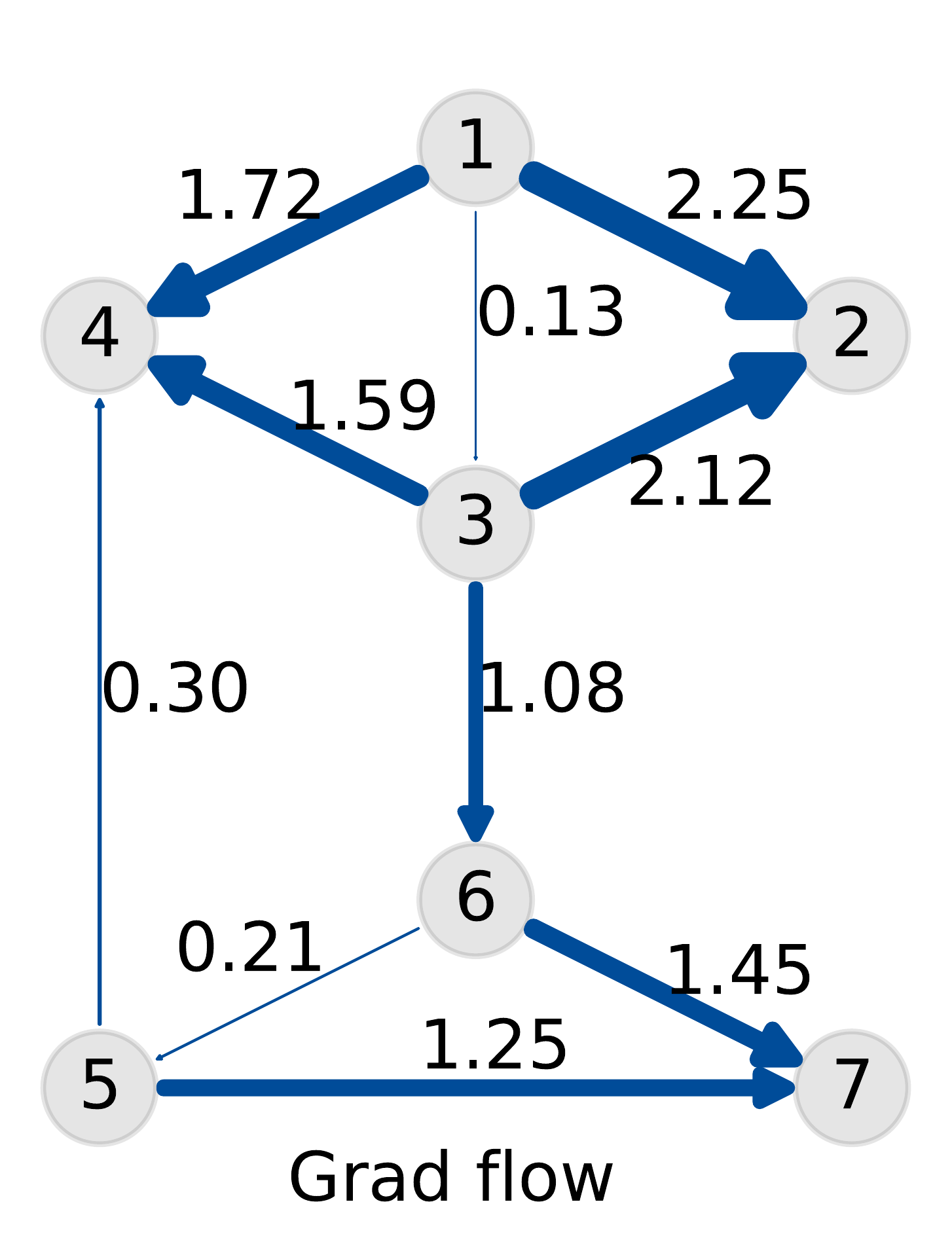}}
  \subfloat[Noisy flow $\bbf$\label{2b}]{%
        \includegraphics[width=0.33\linewidth]{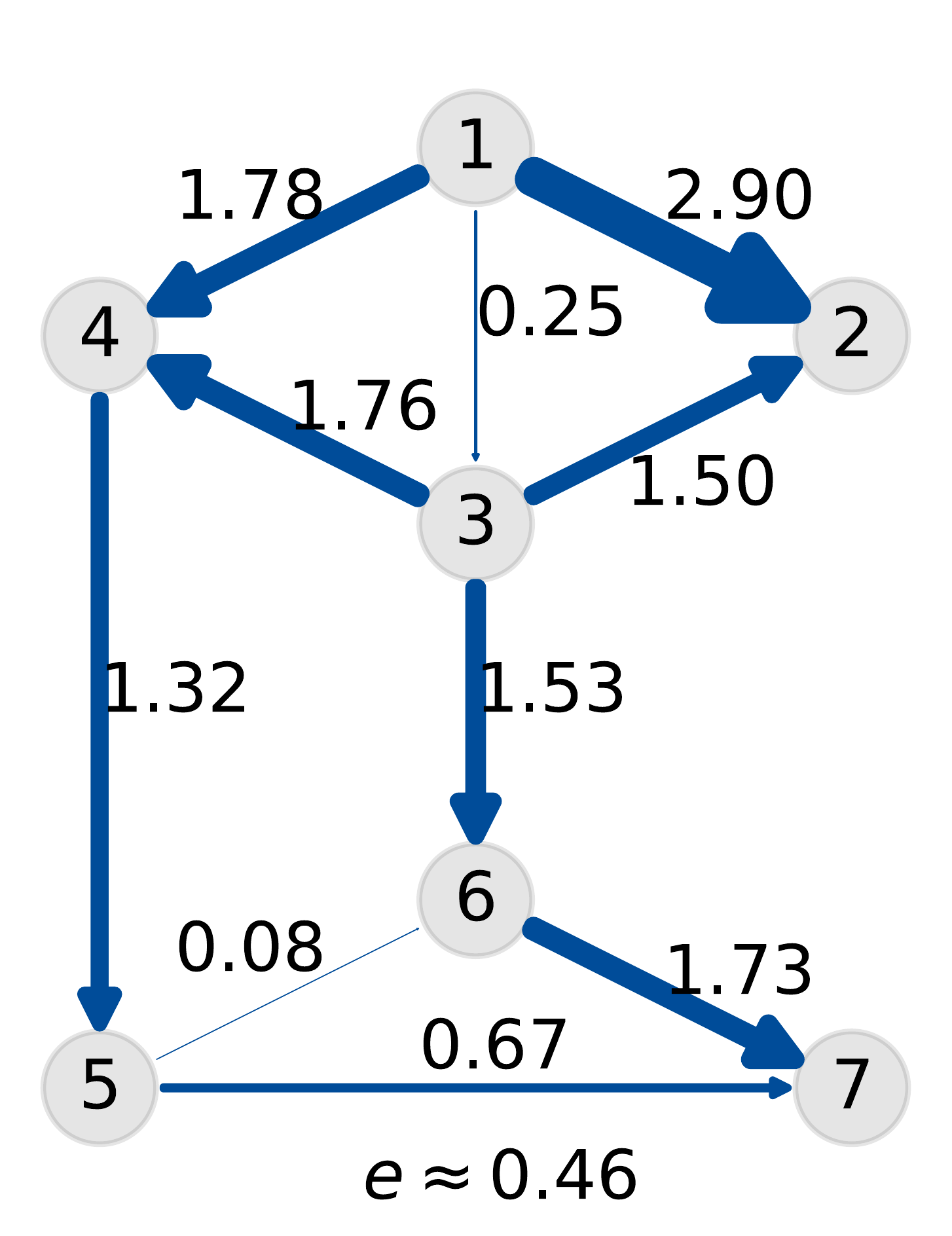}}
  \subfloat[$\bbH = (\bbI+0.5\bbL_1)^{-1}$\label{2c}]{%
        \includegraphics[width=0.33\linewidth]{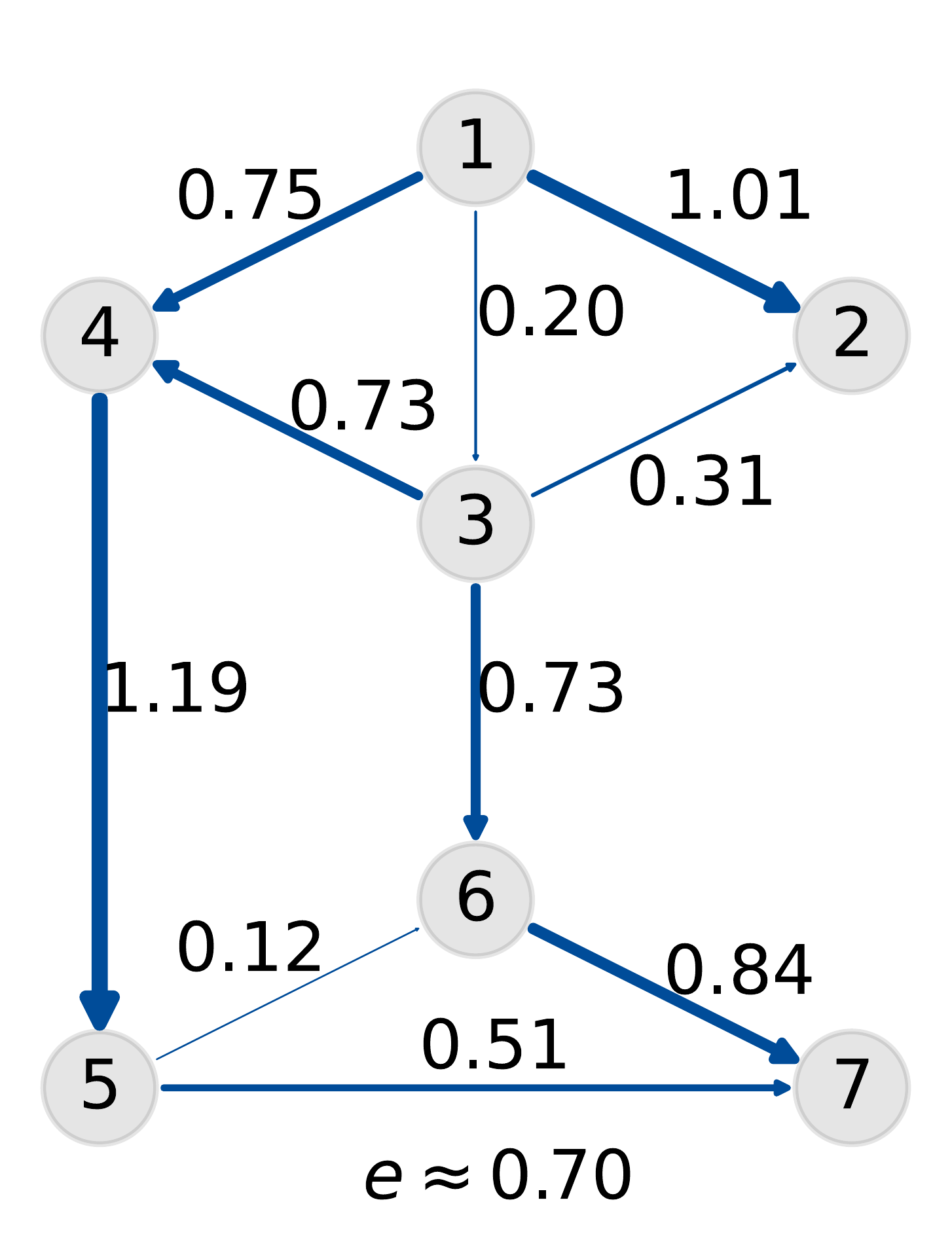}}
  \vspace{-4mm}
  \subfloat[$\bbH = (\bbI+0.5\bbL_{1,\ell})^{-1}$\label{2d}]{%
       \includegraphics[width=0.34\linewidth]{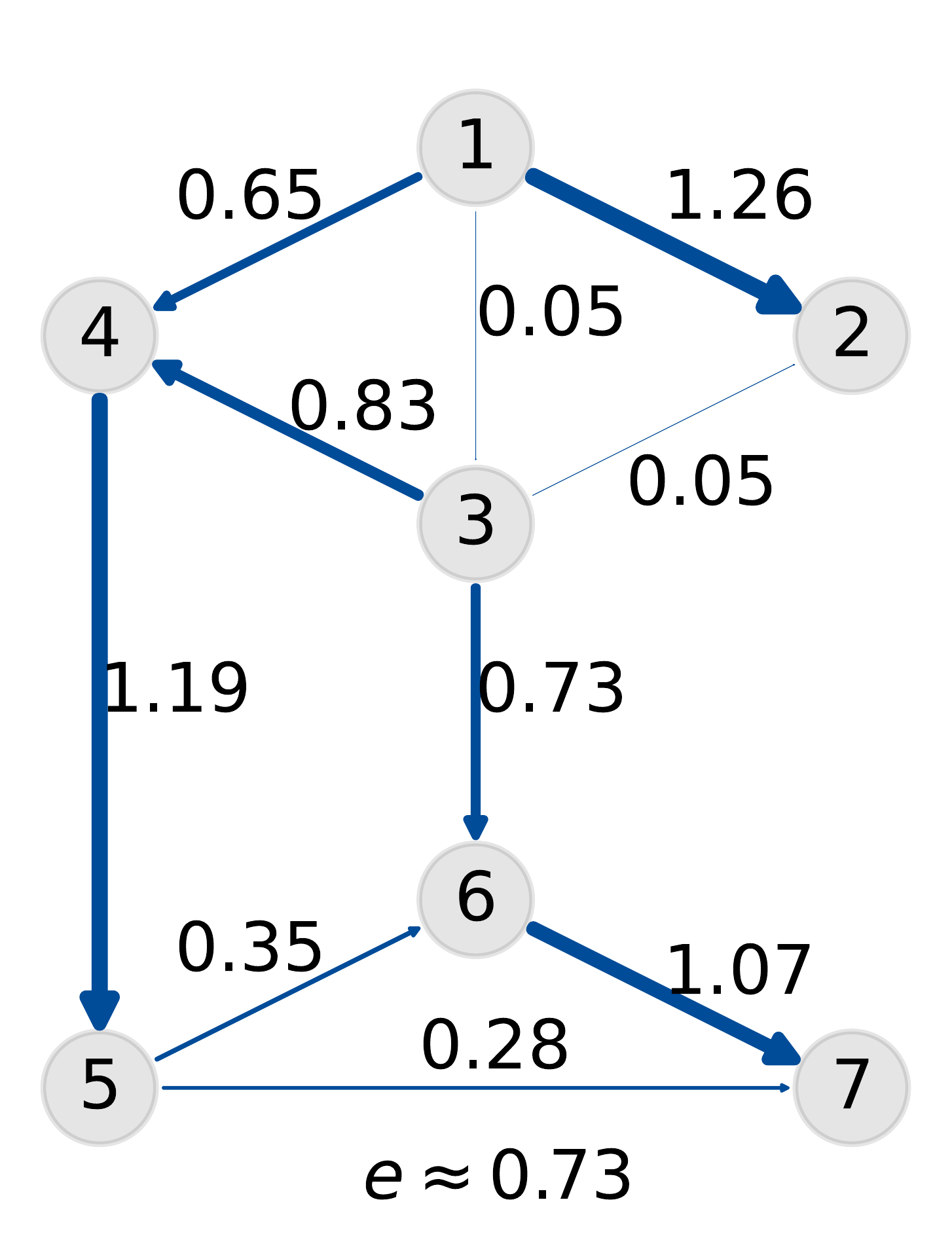}}
  \subfloat[$\bbH_1$, $L=4$\label{2e}]{%
        \includegraphics[width=0.33\linewidth]{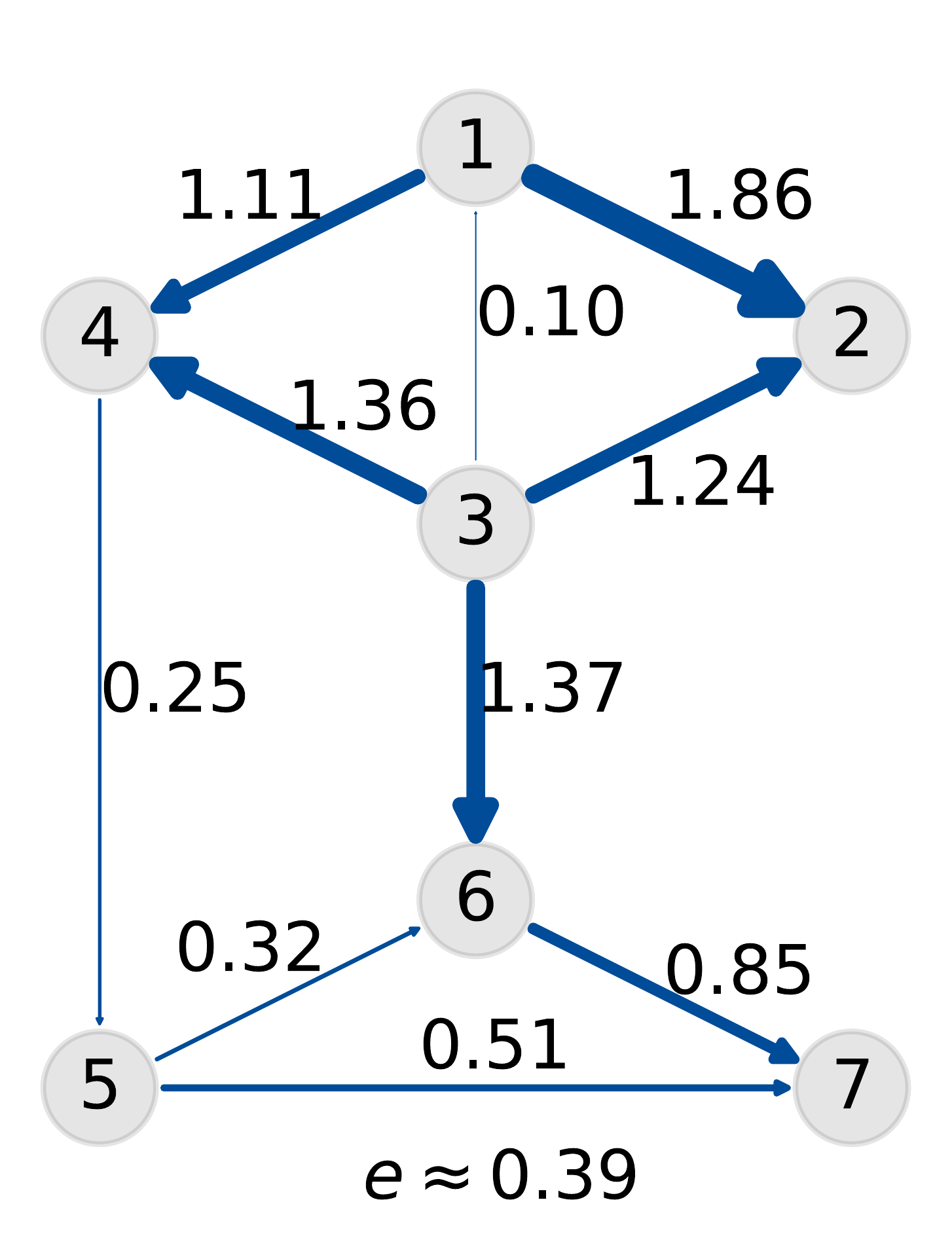}}
  \subfloat[$\bbH_1^{\text{SV}}$, $L_1,L_2=1$\label{2f}]{%
        \includegraphics[width=0.33\linewidth]{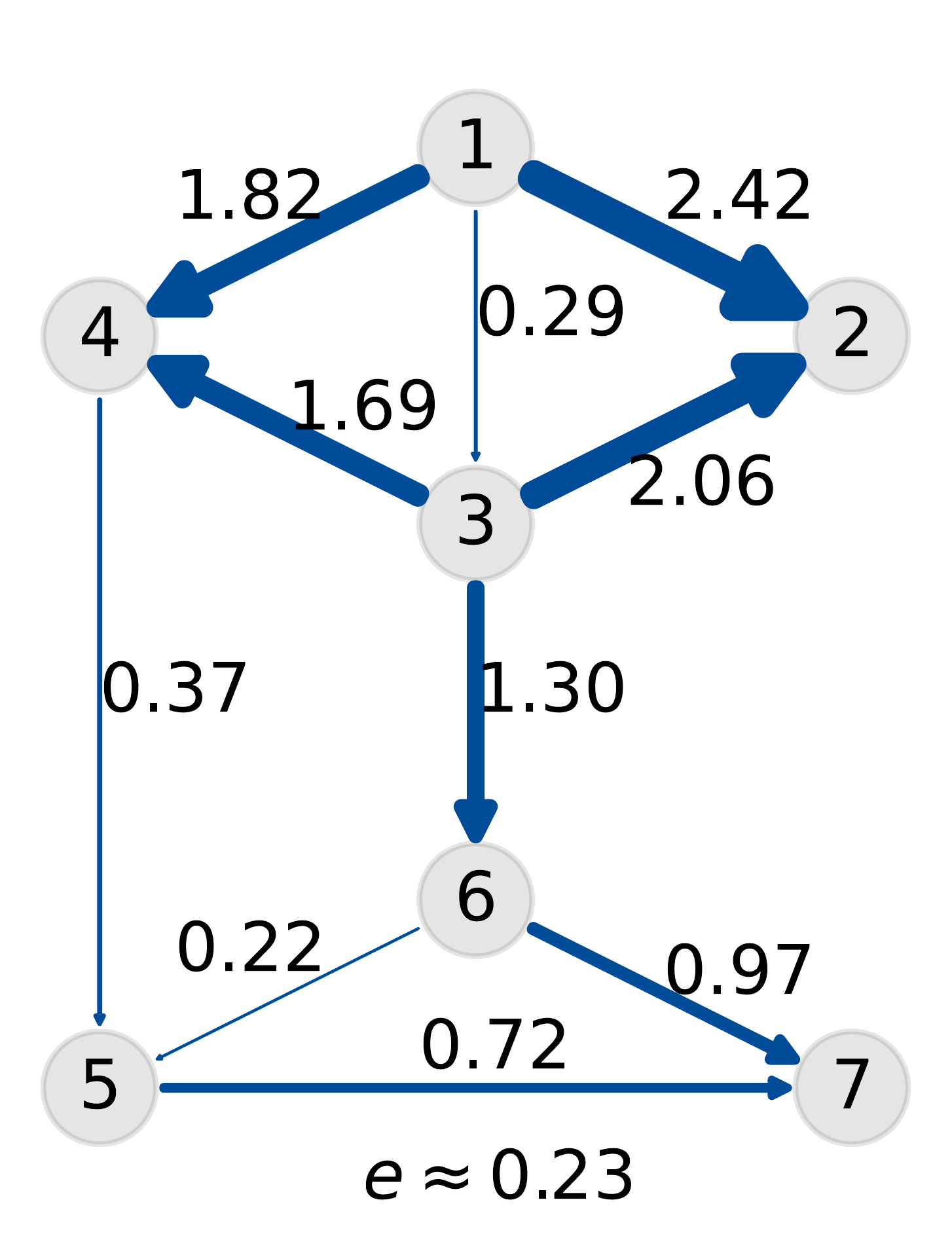}}
  \caption{Gradient flow denoising on a SC (cf. Section \ref{sec:experiments}). The estimation error $e$ is the normalized RMSE. (a) Edge flow $\bbf_0$ induced by a node signal. (b) The noisy observation $\bbf$. (c) Denoised by the filter in \cite{schaub2021}. (d) Denoised by the filter in \cite{schaub2018flow}. (e) Denoised by filter \eqref{eq.filter-1} with length $L=4$. (f) Denoised by the subspace-varying filter \eqref{eq.filter-2} with $L_1=L_2=1$. }
  \label{fig:2} 
  \vspace{-4mm}
\end{figure}

\vspace{1mm}\noindent\textbf{Example 1. } Consider an edge flow containing mainly gradient embeddings with high frequency, i.e., induced by a potential on the nodes (Fig. \ref{2a}). If this edge flow is corrupted by white noise, the filtering proposed in \cite{schaub2018flow}, will not provide a good estimate (Fig. \ref{2d}) of the true flow as it penalizes both high gradient and curl frequencies. Similar to the filter  (Fig. \ref{2c}) regularized by $\bbf^\top\bbL_1\bbf$ proposed in \cite{schaub2021}.
We will see that our simplicial filters can solve such problem.

\vspace{1mm}\noindent\textbf{Remark.} In GSP, graph frequencies measure the signal smoothness over the graph, while simplicial frequencies measure two types of signal properties, which make filtering in the spectral domain more involved. 
For $k=1$, the gradient frequency measures the nodal property, i.e., the net flow passing through the nodes, while the curl frequency measures the rotational property, i.e., the flow circulating along the triangles. The harmonic (zero) frequencies indicate the solenoidal and irrotational properties, i.e., the flow conservation \cite{barbarossa2020,schaub2021}. 
Specifically, observe that a gradient frequency can be larger or smaller than a curl frequency as both frequencies correspond to different, orthogonal features of an edge flow. 
Thus, a low- or high-pass filtering without reference to these type of frequencies thus mix different features of an edge-flow signal.
Filters in simplicial signal spaces thus generally would be expected to tune gradient, curl and harmonic components as required by users.

\section{Simplicial FIR Filter} \label{sec:simplicial filter}
As alluded to above, a linear filter of a flow signal amounts to a map that modulates the gradient, curl and harmonic embeddings of any flow. 
Such a filter may, e.g., extract the gradient component, or remove the harmonic component of a flow. 
While the filter will be dependent on the simplicial complex, it should be independent of the simplicial signals it is applied to.
To achieve this, we propose the following linear simplicial FIR filter:
\begin{equation} \label{eq.filter-1}
    \bbH_1 = \sum_{l=0}^{L-1}h_l\bbL_1^l = \sum_{l=0}^{L-1}h_l (\bbB_1^\top\bbB_1 + \bbB_2 \bbB_2^\top)^l,
\end{equation}
where $L$ is the \emph{filter length},
and $\bbh = [h_0,\dots,h_{L-1}]^\top$ stacks the \emph{filter coefficients}. 
Note that the filter has an analogous form to graph filters \cite{sandryhaila2013discrete,sandryhaila2014discrete}, and can be extended to the $k$-simplicial space by using $\bbL_k$. 
To understand the effect of applying the filter~\eqref{eq.filter-1} to an input signal, we first define the neighborhood set for $k$-simplices.

\vspace{1mm}\noindent\textbf{Simplicial neighborhood.}  
For the $i$-th $k$-simplex $\ccalS_i^k$ in a SC, we define its \emph{lower neighborhood}  $\ccalN_{\ell,i}$ as the set of $k$-simplices sharing a common face with it. Similarly, the \emph{upper neighborhood} $\ccalN_{u,i}$ of $\ccalS_i^k$ collects the $k$-simplices sharing a common coface with $\ccalS_i^k$. 
The lower and upper neighbors of $\ccalS_i^k$ are encoded in the nonzero off-diagonal elements of $\bbL_{1,\ell}$ and $\bbL_{1,u}$, respectively.
More specifically, the cardinality of $\ccalN_{\ell,i}$ (called \emph{lower degree}) and the cardinality of $\ccalN_{u,i}$ (called \emph{upper degree}), are equal to the numbers of nonzero off-diagonal elements of the $i$-th row/column of $\bbL_{1,\ell}$ and $\bbL_{1,u}$, respectively. 
For example, for edge $(5,6)$ in Fig. \ref{1a}, we can see that $\ccalN_{\ell}=\{(4,5),(3,6),(5,7),(6,7)\}$, while the upper neighborhood is $ \ccalN_{u} = \{(5,7),(6,7)\}$, as they share a triangle with edge $(5,6)$. 
The corresponding row/column in $\bbL_{1,\ell}$ and $\bbL_{1,u}$ have respectively $4$ and $2$ nonzero off-diagonal elements.  Note that these local lower and upper connections lead to the sparsity of the Hodge Laplacian.

\begin{comment}
The off-diagonal entry $(i,j)$ of $\bbL_{k,l}$ has values
\[
    \begin{cases}
    1, \text{if the $i$th and $j$th  $k$-simplex has the same orientation} \\
    -1, \text{if the $i$th and $j$th  $k$-simplex has the opposite orientation} \\
    0, \text{otherwise}.
    \end{cases}
\]
The same applies to the off-diagonal entries of $\bbL_{k,u}$.
\end{comment}

\vspace{1mm}\noindent\textbf{Simplicial signal shift.} Applying the filter \eqref{eq.filter-1} to an input signal $\bbf$, i.e., $\bbH_1\bbf$ can be understood in terms of the basic signal shift operation $\bbL_1\bbf$. 
We denote a shift of an edge flow $\bbf$ as
$
    \bbf^{(1)} \triangleq \bbL_1 \bbf = \bbB_1^\top \bbB_1 \bbf + \bbB_2\bbB_2^\top \bbf,
$
where we define $\bbf^{(1)}_{\ell} \triangleq \bbB_1^\top \bbB_1 \bbf$ and $\bbf^{(1)}_{u} \triangleq \bbB_2\bbB_2^\top \bbf$. 
The shifting result on the $i$-th edge, i.e., $[\bbf^{(1)}]_i$, can be expressed as %\mts{writing below a bit more compact (note the union in the sum), old version commented out}
\begin{equation} \label{eq.shifting-per-edge}
    \begin{aligned}
    [\bbf^{(1)}]_i &= [\bbf^{(1)}_{\ell}]_i + [\bbf^{(1)}_{u}]_i\\
    &= \sum_{j\in\{\ccalN_{\ell,i} \cup \;i\}} [\bbL_{1,\ell}]_{ij}[\bbf]_j + \sum_{j\in \{\ccalN_{u,i} \cup \;i \} }[\bbL_{1,u}]_{ij}[\bbf]_j.
\end{aligned}
\end{equation}
Thus, the shift operation in the edge space is local.
For each edge, it collects information from its lower and upper adjacent neighbors and combines this information with its own state. 
Figs.~\ref{1b} and \ref{1c} show the operation of shifting an all-one flow by $\bbL_{\ell,1}$ and $\bbL_{1,u}$ on one edge.
For a given input $\bbf$, we can now write the filter output as
\begin{equation} \label{eq.filter-output-expression}
    \bbf_o = \bbH_1 \bbf = \sum_{l=0}^{L-1} h_l\bbL_1^l \bbf = \sum_{l=0}^{L-1} h_l \bbf^{(l)},
\end{equation}
where $\bbf^{(l)}\triangleq\bbL_1^l\bbf = \bbL_1 \bbf^{(l-1)}$ is the $l$-th shift of the edge flow, which contains information up to its $l$-th hop lower and upper neighbors. 
Thus, the filter output can be seen as a weighted linear combination of consecutively shifted edge flows, which is similar to the concepts of graph signal shifting in graph filters~\cite{sandryhaila2013discrete,sandryhaila2014discrete}. 

\vspace{1mm}\noindent\textbf{Distributed implementation and complexity.} Observe that~\eqref{eq.filter-output-expression} is characterized by the recursion $\bbf^{(l)} = \bbL_1\bbf^{(l-1)}$. That is, edges can locally compute $\bbf^{(l)}$ by exchanging information  with their neighbors about the previous shifted signal, $\bbf^{(l-1)}$.  This implies that in \eqref{eq.filter-output-expression} we need in total $L$ such shifts by $\bbL_1$. 
Moreover, as each shift~\eqref{eq.shifting-per-edge} can be implemented via local operations, which can be implemented by distributed communication between edges, the final output of the simplicial FIR filter can be computed in a distributed way. 
If we denote the maximal edge degree by $D$, then the communication cost to compute~\eqref{eq.shifting-per-edge} if  $\ccalO(D)$. Thus,~\eqref{eq.filter-output-expression} has a total cost of $\ccalO(LN_1D)$. Usually, the edge degree is independent of (and much smaller than) the number of edges, so we would have a complexity of $\ccalO(LN_1)$.

\vspace{1mm}\noindent\textbf{Spectral analysis.}
By applying an eigen-decomposition on the filter \eqref{eq.filter-1}, we obtain
$ %\label{eq.filter-evd}
    \bbH_1 = \sum_{l=0}^{L-1}h_l \bbU_1 \bLambda_1^l \bbU_1^\top = \bbU_1  \Big( \sum_{l=0}^{L-1}h_l \bLambda_1^l \Big) \bbU_1^\top,
$
where we define  $\tilde{\bbH}_1 \triangleq \sum_{l=0}^{L-1}h_l \bLambda_1^l $ as the frequency response of $\bbH_1$. 
For simplicity of notation, let us assume hereafter that $\bLambda_1 = \diag(\lambda_1,\dots,\lambda_{N_1})$. 
At frequency $\lambda_i$, the filter implements the response
$\tilde{H}_1(\lambda_i) = \sum_{l=0}^{L-1}h_l \lambda_i^l$ leading to the spectral input-output relation $\tilde{f}_o(\lambda_i) = \tilde{H}_1(\lambda_i)\tilde{f}(\lambda_i)$. 
As different types of frequencies are present in simplicial signals, filter \eqref{eq.filter-1} needs to control the different signal components according to their corresponding frequencies. 
In other words, the gradient component needs to be tuned using $\tilde{H}_1(\lambda_{i})$ with $\lambda_i \in \ccalQ_G$, the curl component using $\tilde{H}_1(\lambda_{i})$ with $\lambda_i \in \ccalQ_C$, and the harmonic component using $\tilde{H}_1(\lambda_{i})$ with $\lambda_i \in \ccalQ_H$.
By properly designing the frequency response at each frequency, we can thus implement any desired filtering operation.

\vspace{1mm}\noindent\textbf{Filter design.} 
Given a filter specification in the frequency domain $\tilde{H}_1(\lambda_i) =\sum_{l=0}^{L-1}h_l \lambda_i^l= g_i$, for $i=1,\dots,N_1$, we can design the filter coefficients by solving the least squares problem
\begin{equation} \label{eq.filter-design-least-squares}
    \min_{\bbh} \| \bPhi \bbh - \bbg \|^2 ,
\end{equation}
where $\bPhi\in\setR^{N_1\times L}$ is a Vandermonde matrix with entries $[\bPhi]_{i,j} = \lambda_i^{j-1}$, and $\bbg = [g_1, \dots, g_{N_1}]^\top$. 
Hence, given a desired frequency response $\bbg$, we first identify the frequency sets $\ccalQ_G$, $\ccalQ_C$ and $\ccalQ_H$ by computing the eigenvalues
of $\bbL_{1,\ell}$ and $\bbL_{1,u}$, and then solve problem \eqref{eq.filter-design-least-squares} to obtain $\bbh$.
We can then build a simplicial FIR filter which has (approximately) the desired frequency response and apply the filter directly in the simplicial space as in \eqref{eq.filter-output-expression}. 
For instance, to solve the problem in Example 1, we can design a \emph{gradient-preserving} filter by setting the desired frequency response to one at the gradient frequencies and zero at the harmonic and curl frequencies.

\vspace{1mm}\noindent\textbf{Subspace-varying simplicial filter. } 
As seen in Section \ref{sec:fourier transform}, the gradient space is fully described by $\bbL_{1,\ell}$ and the curl space by $\bbL_{1,u}$. 
However, for the filter \eqref{eq.filter-1}, the shift operators $\bbL_{1,\ell}$ and $\bbL_{1,u}$ are jointly weighted by $\{h_l\}_{l=0}^{L-1}$, and consequently we cannot control the frequency response at the gradient and curl frequencies independently.
To enable an independent design of the frequency responses at gradient and curl frequencies we thus assign $\bbL_{1,\ell}$ and $\bbL_{1,u}$ a different set of weights, leading to the following \emph{subspace-varying FIR filter}
\begin{equation} \label{eq.filter-2}
    \bbH_1^{\text{SV}} = h_0\bbI + \sum_{l_1=1}^{L_1} \alpha_{l_1} (\bbB_1^\top\bbB_1)^{l_1} + \sum_{l_2=1}^{L_2} \beta_{l_2} (\bbB_2 \bbB_2^\top)^{l_2},
\end{equation}
where $\balpha = [\alpha_1,\dots,\alpha_{L_1}]^\top$ and $\bbeta = [\beta_1,\dots,\beta_{L_2}]^\top$ control the gradient and curl frequencies, and the term $h_0$ controls the harmonic frequency response. 
Here we use the convention that if $L_1=0$ or $L_2=0$ the corresponding terms are zero. 
This can be useful, if we only want to tune the gradient or curl component.
Our earlier discussion on the shift operator and the complexity also apply to \eqref{eq.filter-2}.

Using a spectral decomposition of~\eqref{eq.filter-2}, we can see its frequency response is given by
% By applying an eigen-decomposition on~\eqref{eq.filter-2} (the eigenvectors will be given again by $\bbU_1$), we find its frequency response as
\begin{equation} \label{eq.freq-response-form2}
    \tilde{H}_1^{\text{SV}}(\lambda_i) = \begin{cases}
    h_0, \text{ for } \lambda_i \in \ccalQ_H, \\
    h_0 + \sum_{l_1 = 1}^{L_1}\alpha_{l_1}\lambda_{i}^{l_1}, \text{ for }  \lambda_i \in \ccalQ_G, \\
    h_0 + \sum_{l_2=1}^{L_2}\beta_{l_2}\lambda_{i}^{l_2}, \text{ for }  \lambda_i \in \ccalQ_C. \\
    \end{cases}
\end{equation}
From these equations, we see that in the subspace-varying filter, the frequency responses at gradient and curl frequencies can indeed be controlled independently via the different weights.
As a result, the filter is more flexible than $\bbH_1$ (and has increased degrees of freedom).

For a specified frequency response $\tilde{\bbH}_1^{\text{SV}} = \diag(\bbg)$, we can again design the filter coefficients by solving the least squares problem
\begin{equation}\label{eq.filter-design-form2}
    \min_{h_0,\balpha,\bbeta} \left\|  \begin{bmatrix} {\bf 1} \ \vline &  \begin{matrix} {\bf 0}^\top \\ \hline \begin{matrix} 
    \bPhi_G & {\bf 0}_{N_G \times L_2} \\ {\bf 0}_{N_C \times L_1} & \bPhi_C
    \end{matrix} \end{matrix}  \end{bmatrix} 
    \begin{bmatrix}
    h_0 \\ \balpha \\ \bbeta
    \end{bmatrix}
    - \bbg
    \right\|,
\end{equation}
where ${\bf 1}$ (${\bf 0}$) is an all-one (all-zero) vector, ${\bf 0}_{m \times n}$ is an all-zero matrix of size $m \times n$, and $\bPhi_G\in\setR^{ N_G  \times L_1}$ and $\bPhi_C\in\setR^{ N_C \times L_2}$ are Vandermonde matrices with respective entries $[\bPhi_G]_{i,j} = \lambda_{G,i}^{j}$ and $[\bPhi_C]_{i,j} = \lambda_{C,i}^{j}$.
An immediate observation is that when $L_1$ (or $L_2$) is zero, the number of rows for this system of equations is one plus the number of curl (or gradient) frequencies, which is less than the corresponding number of $\bPhi$ in~\eqref{eq.filter-design-least-squares}. 
This makes the filter \eqref{eq.filter-2} more suitable when only gradient (or curl) components need to be tuned, as the remaining can be jointly controlled by the frequency response $h_0$ at zero as in \eqref{eq.freq-response-form2}.

\begin{comment}
\begin{figure}[!t]
    \vspace{-9mm}
    \centering
    \includegraphics[width = 2.5in, scale = 1]{figures/gradient_preserving_filter_response.eps}
    \caption{Frequency response of gradient-preserving filters $\bbH_1$ and $\bbH_1^{\text{SV}}$ with a total length of 5 for the edges of the SC in Fig. \ref{1a}. The curl frequencies are $\{2,3,4\}$ and the remaining nonzero eigenvalues are gradient frequencies. Both are designed by solving the related least squares problems. With the same total lengths, the subspace-varying filter does much better than the filter $\bbH_1$, which has large frequency leaking at curl frequencies.
    }
    \label{fig:gradient-preserving-frequency-response}
    \vspace{-6mm}
\end{figure}
\end{comment}

\section{Numerical experiments} \label{sec:experiments}
In this section, we provide experimental results for the simplicial filters \eqref{eq.filter-1} and \eqref{eq.filter-2} to process the edge flow in a network. To evaluate the performance, we consider the normalized rooted mean square error (NRMSE), defined as $e = \lVert \hat{\bbf} - \bbf_0 \lVert_2/ \lVert \bbf_0 \lVert_2$, where $\hat{\bbf}$ is the flow estimate by filters and $\bbf_0$ is the corresponding groundtruth flow.

\begin{figure}[!t]
    \vspace{-4mm}
    \centering
    \includegraphics[width = 2.5in, scale = 1]{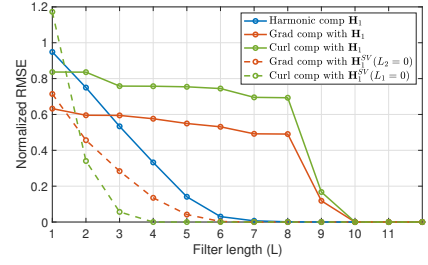}
    \caption{Sub-component extraction by filters $\bbH_1$ and $\bbH_1^{\text{SV}}$. The extraction becomes better as the filter length grows. But for gradient and curl component extraction, the filter $\bbH_1$ cannot obtain an accurate extraction until the filter length is nine. However, $\bbH_1^{\text{SV}}$ performs much better than $\bbH_1$, which is consistent to the discussion after \eqref{eq.filter-design-form2}.}
    \label{fig:subcomponent-extraction-synthetic}
    \vspace{-4mm}
\end{figure}

\vspace{1mm}\noindent\textbf{Sub-component extraction.}
We first constructed a synthetic SC as in Fig. \ref{1a} with seven nodes, ten edges, and three triangles \cite{schaub2021}. 
We then generated an edge flow $\bbf = \bbU_1 \tilde{\bbf}$ with $\tilde{\bbf} = \bb1\in\setR^{10}$. 
We then designed FIR filters to preserve only the gradient, curl or harmonic signal components respectively, and compared their results to an exact projection of the signals on the corresponding subspace (see \cite{schaub2021,barbarossa2020,lim2015hodge,schaub2018flow}) in terms of the error.

Fig.~\ref{fig:subcomponent-extraction-synthetic} displays the performance of filter \eqref{eq.filter-1} (solid line) and the subspace-varying filter \eqref{eq.filter-2} (dashed line). 
As expected, the subspace extraction becomes more accurate as $L$ grows (indeed for a filter of order $L=10$ the least squares problem can always be solved exactly).
Importantly, to extract a particular subspace component, filter \eqref{eq.filter-1} performs always worse than the subspace-varying filter \eqref{eq.filter-2}. 
The reason is that for filter~\eqref{eq.filter-2}, we can tune the frequency response in the respective subspace independently, and thus require less coefficients to approximate the desired frequency response.

\vspace{1mm}\noindent\textbf{Edge flow denoising.} 
Using the same network again, we generated an edge flow induced by a node signal which is all-ones in the graph frequency domain and corrupted by white noise, i.e., similar to an instance of the problem in Example 1. 
The noisy flow (Fig. \ref{2b}) introduces an error of $0.46$ originally.
When using the (low-pass) denoising filters discussed in \cite{schaub2018flow} and \cite{schaub2021} (Figs. \ref{2c} and \ref{2d}), the error becomes even larger. This is expected as the assumption of a low-pass signal used in \cite{schaub2018flow, schaub2021} is not fulfilled in our problem setup. 
Using a proper filter design that preserves the gradient components, filter \eqref{eq.filter-1} with $L=4$ and filter \eqref{eq.filter-2} with $L_1=L_2=1$ can, however, achieve errors of $0.39$ and $0.23$, as shown in Figs. \ref{2e} and \ref{2f}.

\vspace{1mm}\noindent\textbf{Sioux Falls network.} 
We now consider a real world Sioux Falls transportation network \cite{leblanc1975algorithm}. 
It contains 24 nodes, 38 edges, and 2 triangles ($1$-simplices). 
We assumed an autoregressive (AR) model to generate a set of time-evolving edge flows as 
\begin{equation} \label{eq.ar_model}
    \bbf_{t+1} = (0.5\bbI+0.3\bbB_1^\top\bbB_1 + \bbB_2\bbB_2^\top + 0.5(\bbB_2\bbB_2^\top)^2 )^{-1}\bbf_{t},
\end{equation}
where $t$ is the time instance.
Assuming the model to be unknown to the analyst, we sought to train filters \eqref{eq.filter-1} and \eqref{eq.filter-2} in order to learn a data-driven model for flow prediction.
Accordingly, we generated a training set containing 20 sample pairs $\{\bbf_{s,\text{in}},\bbf_{s,\text{out}}\}$, where $\bbf_{s,\text{in}}$ is a random Gaussian flow to be used as input for model \eqref{eq.ar_model}, and $\bbf_{s,\text{out}}$ is the output. Note that we can rewrite the filtering equation \eqref{eq.filter-output-expression} as 
\begin{equation} \label{eq.data_driven_filtering}
     [\bbf,\bbf^{(1)},\dots,\bbf^{(L-1)}]\bbh = \bbf_o,
\end{equation}
where the system matrix contains the shifted versions of the input.
With the training set, we built the system matrices by shifting each input and vertically concatenated them and the outputs. Then, we used the least squares solution to \eqref{eq.data_driven_filtering} as the trained filter coefficients $\bbh$ to build a data-driven filter $\bbH_1$. The test data, $\ccalT = \{\bbf_0,\bbf_1,\dots,\bbf_{79}\}$, is generated by initializing \eqref{eq.ar_model} with a random flow $\bbf_0$. Then, with each observation $\bbf_t$, $t=0,1,\dots,79$, we obtained a one-step prediction $\hat{\bbf}_{t+1} = \bbH_1 \bbf_t$. Finally, we evaluated the performance by computing the error between $\bbf_{t+1}$ and $\hat{\bbf}_{t+1}$. Similar steps were followed for the subspace-varying filter $\bbH_1^{\text{SV}}$.

\begin{table}[!t]
\vspace{-1mm}
\caption{Flow prediction on Sioux Falls network. $L_\text{total}$ is the total filter length and $e_1$ and $e_2$ the prediction errors by $\bbH_1$ and $\bbH_1^{\text{SV}}$ on the test data. }
\centering
\vspace{-1mm}
{\scriptsize
\begin{tabular}{c|cc|c|cc}
\thickhline
$L_\text{total}$ & $e_1$ & $e_2$  & $L_\text{total}$ & $e_1$ & $e_2$ \\ \thickhline
\rowcolor{Gray}1 & 0.794 & -- & 6 & 0.268 & \textbf{0.230} \\ \hline
 2 & 0.687 & \textbf{0.597} & 7 & 0.236 & \textbf{0.187} \\ \hline
\rowcolor{Gray}3 & \textbf{0.482} & 0.569 & 8 & 0.207 & \textbf{0.157} \\ \hline
4 & \textbf{0.379} & 0.395 & 9 & 0.185 & \textbf{0.135} \\ \hline
\rowcolor{Gray}5 & 0.308 & \textbf{0.293} & 10 & 0.167 & \textbf{0.118}  \\ \hline
\thickhline
\end{tabular}}
\label{tab:prediction-siouxfalls}
\vspace{-3mm}
\end{table}

The results are shown in Table. \ref{tab:prediction-siouxfalls}. 
We reported the best performance for each total length for $\bbH_1^{\text{SV}}$. 
When $L_\text{total}$ becomes larger, both filters are able to learn the AR model \eqref{eq.ar_model} better and we observe that the prediction becomes more accurate as well. 
Moreover, to deal with models like \eqref{eq.ar_model} in the simplicial space, the more flexible subspace-varying filter $\bbH_1^{\text{SV}}$ usually is able to learn the model better than $\bbH_1$. 

\section{Conclusion}
We proposed two linear FIR filters in the form of \eqref{eq.filter-1} and \eqref{eq.filter-2} for filtering simplicial signals (and more specifically edge flows). 
We revisited the simplicial Fourier transform, and defined gradient and curl frequencies for edge flows, which measure the divergence and rotational properties of the flow. 
Both of our proposed filters are able to tune signals for these two kinds of frequencies and can be distributively implemented, since the simplicial shift operator defined by the Hodge Laplacian can be built from local operations. 
We showed that the subspace-varying filter is more flexible and is especially beneficial if we desire to tune only gradient  or curl frequencies, since the different weights on the lower and upper Laplacians in \eqref{eq.filter-2} enable an independent tuning of those components.
Finally, experimental results were reported to support our findings.
Future work will concern:
\begin{inparaitem}
    \item[1)] examining the utility of simplicial filters beyond the edge space;
    \item[2)] performing filter design for both \eqref{eq.filter-1} and \eqref{eq.filter-2} with smaller costs;
    \item[3)] combining simplicial filters into simplicial neural networks as considered, e.g., in \cite{roddenberry2019hodgenet}.
\end{inparaitem}

\bibliographystyle{IEEETtran}
\bibliography{refs}
\end{document}